\documentclass[11pt]{article}
\pdfoutput=1

\usepackage[T1]{fontenc}              
\usepackage[latin9]{inputenc}         

\usepackage[a4paper]{geometry}        

\usepackage{url}

\usepackage{booktabs}

\usepackage{tabularx}
\newcolumntype{Y}{>{\raggedleft\arraybackslash}X}

\usepackage{graphicx} 
\usepackage{amssymb,amsfonts,amsmath}

\usepackage{natbib}
\usepackage{mathpazo}
\usepackage{microtype}

\usepackage{color}

\newcommand\bI{\boldsymbol I}

\newcommand\bmu{\boldsymbol \mu}

\newcommand\bz{\boldsymbol z}

\newcommand\prob{\text{Pr}}
\newcommand\expect{\text{E}}

\newcommand\var{\text{Var}}

\newcommand\fdr{\text{fdr}}
\newcommand\fndr{\text{fndr}}
\newcommand\Fdr{\text{Fdr}}

\newcommand\CB{\text{CB}}
\newcommand\KS{\text{KS}}
\newcommand\HC{\text{HC}}

\newcommand{\figcite}[1]{Fig.~\textbf{\ref{#1}}}
\newcommand{\eqcite}[1]{Eq.~\textbf{\ref{#1}}}
\newcommand{\tabcite}[1]{Tab.~\textbf{\ref{#1}}}

\begin{document}

\date{12 December 2011; last revised 19 July 2012}

\title{
Signal Identification for Rare and Weak Features: Higher Criticism or False Discovery Rates?
}

\author{Bernd Klaus 
      \thanks{Institute for Medical Informatics,
      Statistics and Epidemiology,
      University of Leipzig,
      H\"artelstr. 16--18,
      D-04107 Leipzig, Germany} 
{} and
 Korbinian Strimmer \footnotemark[1]
}

\maketitle
\begin{abstract}

Signal identification in large-dimensional settings is a challenging problem
in biostatistics.  Recently, the method of higher criticism (HC) was shown to
be an effective means for determining appropriate decision thresholds.
 Here, we study HC  from a false discovery rate (FDR) perspective.
We show that the HC threshold may be viewed as an approximation
to a natural class boundary (CB) in two-class discriminant analysis
 which in turn is expressible as FDR threshold.
We demonstrate that in a rare-weak setting in the 
region of the phase space where signal identification is
possible  both thresholds are practicably indistinguishable,
and thus HC thresholding is identical to using a simple local FDR cutoff.
The relationship  of the HC and  CB thresholds and their properties 
are investigated  both analytically and by simulations, and are further
compared  by application to four cancer gene expression data sets.

\end{abstract}

\newpage

\section{Introduction}

Identification of sparse and weak signals in complex high-dimensional data 
is a challenging  statistical problem that has many important applications 
in fields as diverse as astronomy, finance, genetics, medicine, and proteomics.
A typical biomedical task is the search for biomarkers using data from
genome-wide association studies \citep{XCL2011}.
Signal \emph{identification} is much more difficult than the closely related
problem of signal \emph{detection}.  Whereas in detection we are concerned 
purely with the presence or absence of a signal,  in identification  
we additionally seek to locate the signal.

In a series of recent publications the method of ``Higher Criticism'' (HC) 
was powerfully advocated in settings with rare and weak features as an 
efficient means for signal detection \citep{DJ2004} as well as signal
identification \citep{DJ08,DJ2009}.  Originally, HC was introduced by 
\citet{Tukey1976} as an approach to multiple significance testing using 
a second level test statistic computed from $p$-values.  Importantly, in 
\citet{DJ2004}  it was shown that HC provides a procedure that is optimal 
for signal detection in the sense that it achieves the best possible 
theoretical detection limit discovered earlier by \citet{Ing1999}.  
Subsequently, HC was also employed in a thresholding procedure to determine
relevant features for prediction. Again, it was demonstrated that
the HC approach to signal identification outperforms other commonly employed
selection strategies, in particular those based on false discovery rates 
\citep{DJ08,DJ2009}. 

In \citet{AS2010} the utility of HC for variable selection in classification
was confirmed but at the same time it was also empirically shown that in the signal 
identification problem controlling the false \emph{non}-discovery rate is
equivalent to the HC procedure. 
Furthermore, it was discovered by \citet{JW2007} that HC is not unique in 
achieving the detection limit.  Given the success of HC this raises questions
about the fundamental principles that may underlie this approach.

Here, we explore signal identification using the HC and false (non)-discovery 
approaches, with the aim to provide a better understanding of HC as well 
as offering a simple explanation for HC's favorable performance.  
Specifically, we argue that the decision threshold provided by HC
may also be viewed as an approximation 
to a natural class boundary (CB) in classification which in turn is easy to understand from 
a false discovery rate perspective. In particular, in the rare-weak setting 
in the region of the phase space where identification  is actually possible 
we show that the HC and CB threshold are nearly indistinguishable.

The remainder of the paper is structured as follows.
First, we provide a  non-technical introduction to HC both on sample and 
population level.
Second, we derive the ideal thresholds corresponding to HC and false discovery 
rate approaches, and explore their mutual relationships.
Next, we investigate these thresholds in the rare-weak model and establish the
near identity of HC and a natural CB threshold in the rare-weak 
identification setting. Finally, we demonstrate the validity of the
theoretical considerations
by simulation and by analyzing data from four gene expression experiments.

\section{Higher Criticism}

In the following, we introduce the HC approach to
signal identification, and discuss various properties of the HC threshold
both from a sample and population point of view.

\subsection{Empirical HC threshold based on $p$-values}

We consider a situation with $d$ observed test statistics $y_1, \ldots, y_d$.
For each statistic we compute a corresponding  $p$-value $p_1, \ldots, p_d$.
The dimension $d$ is potentially very large, as in many current applications
in genomics or proteomics.

The HC approach to signal identification then proceeds 
as follows: 
\begin{itemize}
\item
First, by arranging the $p$-values from smallest to largest
$p_{(1)}, \ldots, p_{(d)}$  the
empirical distribution function of the $p$-values is obtained,
$$
\hat F(x) = i/d \,\,\text{for}\,\, p_{(i)} \leq x < p_{(i+1)}
$$
with $x \in [0; 1]$, $p_{(0)}=0$ and $p_{(d+1)}=1$.
\item
Second, the
 empirical HC objective function 
\begin{equation}
\widehat{HC}(x ) =  { | \hat F( x )  -  x | \over  \sqrt{ \hat F(x ) (1- \hat F(x) ) /d } } \,
\label{eq:hcfunc}
\end{equation}
 is computed \citep{DJ08,DJ2009}. 
\item
Third,  the HC statistic $\widehat{HC}^\star$ is obtained as the maximum
of the empirical HC objective
$$
\widehat{HC}^\star = \max_{i}  \widehat{HC}(p_{(i)} ) = \widehat{HC}(x^{\HC}) \,.
$$
\item Finally, the maximizing argument $x^{\HC}$
is taken as the HC decision threshold for signal identification.
As shown in \figcite{fig:thresh}{\bf a}
all  $p_i < x^{\HC}$
are considered  ``significant''
and thus assumed to likely correspond to non-null cases.

\end{itemize}

\begin{figure}[thp!]
\begin{center}
\centerline{\includegraphics[width=.9\textwidth]{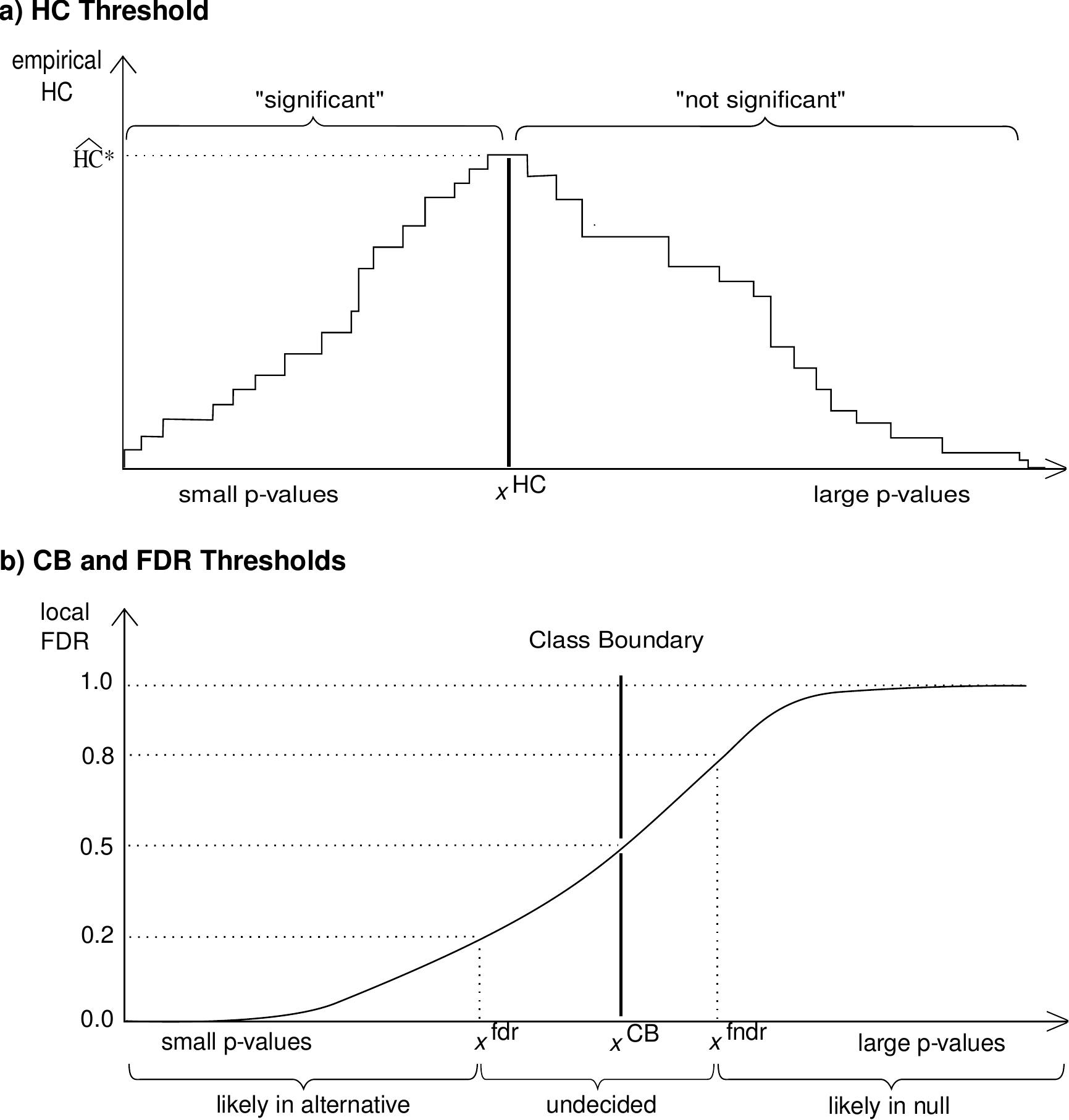}}
\caption{a) Empirical HC decision threshold $x^{\HC}$ obtained by maximizing 
the empirical HC objective, and
 b) Class boundary $x^{\CB}$ given by local FDR = 1/2
and its relationship to the neighboring local FDR and local FNDR thresholds.
\label{fig:thresh}}
\end{center}
\end{figure}

Informally, the empirical HC objective function $\widehat{HC}(x )$ may 
be interpreted as
$z$-scores constructed from $p$-values --- recall that 
$\var(\hat{F}(x)) = F(x) (1-F(x)) / d$.   Indeed, it is precisely
this second level
assessment of $p$-values that was the original motivation for the HC  
approach \citep{Tukey1976} and that gave rise to its name  ``Higher Criticism''.

\subsection{Population HC objective function 
and goodness-of-fit statistics}

By definition, $p$-values have a uniform $U(0,1)$ null distribution 
with $ F_0( x ) =x$.  Moreover, the marginal distribution of the $p$-values
may be viewed as a two-component mixture 
$$
F( x ) = \eta_0 F_0(x) + (1-\eta_0) F_A(x)
$$
 of the null model $F_0(x)$ and an alternative model $F_A(x)$
 where $\eta_0 \in [0;1]$ is the proportion of the null.
With this in mind the squared empirical HC objective function can be written 
as 
$$
\widehat{HC}(x )^2 \propto { ( \hat F_A( x )  -  F_0(x) )^2 \over  \hat F(x ) (1- \hat F( x) )  } \,.
$$
The proportionality factor $d (1-\eta_0)^2$ has been left out as it does not depend on $x$
and hence is irrelevant for determining the
decision threshold $x^{HC}$.  Thus, for maximization we
can use the above formula rather than \eqcite{eq:hcfunc}. Furthermore,
it has the advantage of immediately generalizing to the \emph{population level}
(i.e. to $d \rightarrow \infty$)
\begin{equation}
{HC}(x )^2 \propto { (  F_A(x )  -  F_0(x) )^2 \over  {  F(x ) (1-  F( x) ) } } 
\label{eq:hc2-popfunc}
\end{equation}
which greatly facilitates the conceptual understanding of the HC approach. 

The function \eqcite{eq:hc2-popfunc} is well known from the goodness-of-fit statistic 
of \citet{AndersonDarling1954} which is proportional to
 the expectation $\expect_F(HC(X)^2)$.
Hence, 
the HC statistic bears the same relationship to the Anderson-Darling statistic
as does the Kolmogorov-Smirnov statistic to the Cram\'er-van Mises statistic
\citep{Darling1957}.  Moreover, as can be seen in  \tabcite{tab:goodnessoffit}
 the HC statistic is the
standardized Kolmogorov-Smirnov (KS) statistic. In fact, the KS statistic
may used in the same fashion as HC to derive a decision threshold 
$x^{KS}$.


\begin{table}[t]
\caption{Relationship of HC statistic with other goodness-of-fit statistics.}
\begin{center}
\begin{tabular}{lcc}
\toprule
                 & Supremum & Expectation \\
\midrule
\midrule
Not standardized & Kolmogorov-Smirnov: & Cram\'er-von Mises: \\
                 & $\sup_x |F_A(x)-F_0(x)|$ 
                 & $\expect_F \{  (F_A(X)-F_0(X))^2  \}$\\
\\
Standardized     & Higher Criticism: & Anderson-Darling: \\
    & $\sup_x \left\{ { |F_A(x)-F_0(x)| \over \sqrt{F(x) ( 1-F(x))}} \right\}$
    & $\expect_F \left\{ { (F_A(X)-F_0(X))^2 \over F(X) ( 1-F(X))} \right\}$ \\
\bottomrule 
\end{tabular}
\end{center}
\label{tab:goodnessoffit}
\end{table}


In the mixture model for $p$-values it is commonly assumed (see also Section~3
on false discovery rates) that $F_A(x) \geq F_0(x)$ for all $x$, i.e.
that the
the alternative component is stochastically smaller than or equal to the
null component.  Thus, on population level
(though not on sample level) we may leave out the absolute
value signs in the first column of \tabcite{tab:goodnessoffit}.

\subsection{Invariance of HC objective function}

By inspection of \eqcite{eq:hc2-popfunc} we can derive a number of interesting properties
 of the HC objective function.

First, it is completely symmetric with regard to the two
components in the underlying mixture model for the $p$-values.
The alternative model $F_A$ and  the null model $F_0$ play 
the same role in~\eqcite{eq:hc2-popfunc}. 

Second, for computing the HC
objective it is not necessary to explicitly 
specify the null proportion $\eta_0$.

Third, \eqcite{eq:hc2-popfunc} is invariant against transformation
of the underlying test statistic. This can be seen as follows: 
Under a change of variables from $x$ to $y = y(x)$
the distribution function changes according to
$$
F^Y(y) =  
\begin{cases}
F\biggl( x(y) \biggr)    & \text{for increasing $x(y)$, and} \\
1 - F\biggl( x(y) \biggr)    & \text{for a decreasing transformation.} \\
\end{cases}
$$
Applied to \eqcite{eq:hc2-popfunc} this leads to
$$
{HC}(y )^2 = {HC}(y(x) )^2 \propto { (  F^Y_A(y )  -  F^Y_0(y) )^2 \over  {  F^Y(y ) (1-  F^Y( y) ) } } \, .
$$
Remarkably, the HC objective function  \eqcite{eq:hc2-popfunc}
retains its functional form under a change of variables.
Thus, \eqcite{eq:hc2-popfunc} is \emph{not} constrained
to $p$-values only and may instead  be applied to any 
test statistic $y$ without the need of prior conversion to
the $p$-value scale.  The HC decision threshold as the location of the
maximum of \eqcite{eq:hc2-popfunc}  transforms accordingly, 
from $x^{\HC}$ to $y^{\HC} =  y(x^{\HC})$.

\section{False Discovery Rates}

For comparison we now briefly recapitulate the ``False Discovery Rate'' (FDR) approach 
to signal identification.  Like the HC approach it is also best understood on
the population level.  For comprehensive overview see, e.g., \citet{Efr08a}.

\subsection{Definition of FDR and related quantities}

Essentially, there are two variants of FDR criteria, one based on distributions (tail area-based FDR)
and the other on densities (local FDR).  In addition, if the roles of null and alternative
are interchanged one arrives at the ``False Non-Discovery Rate'' (FNDR).

On the $p$-value scale, the tail-area-based FDR (or Fdr)
is defined as
$$
\Fdr(x) = \prob(\text{``null''} | X \leq x) = {\eta_0 F_0(x) \over F(x)} = {\eta_0 x \over F(x)} \,.
$$
By construction, $\Fdr(x)$ is the proportion of $p$-values from the null component 
found among all $p$-values smaller than $x$.
In order for $\Fdr(x)$ to be monotonically increasing with $x$ 
(i.e. to ensure that the ordering of test statistics does not change)
it is
necessary that $f_A(x)$ is a monotonically decreasing density, and thus
both $F_A(x)$ and $F(x)$ must be assumed to be concave \citep{LLF05,Str08c}.  
This also implies that
the alternative and null are stochastically ordered with 
$F_A(x) \geq F_0(x)$ for all $x$.
The empirical estimate of Fdr for a set ordered $p$-values
$p_{(1)}, \ldots, p_{(d)}$ is the rule of \citet{BH95},
$$
\widehat{\Fdr}(p_{(i)}) = {\hat\eta_0 p_{(i)} \over \hat{F}(p_{(i)})} \leq \frac{ d  }{i} p_{(i)}\, ,
$$
which also shows that
Fdr may be viewed as a multiplicity-adjusted $p$-value.
As complementary error one also studies the tail-area based FNDR 
 that is the proportion of non-null $p$-values among 
$p$-values larger than $x$. On the $p$-value
scale it is defined as
$$
\text{Fndr}(x) =  \prob(\text{``alternative''} | X \geq x) = (1-\eta_0) {1- F_A(x) \over 1- F(x)} \,.
$$
Fndr and Fdr play a similar role
as sensitivity and specificity  in classical testing \citep{GW02}.

Local FDR (fdr) is a density-based quantity defined as
the probability of the null under the observed data,
\begin{equation}
\fdr(x)  = \prob(\text{``null''} | X = x) = {\eta_0 f_0(x) \over f(x)} = {\eta_0 \over f(x)}\, 
\label{eq:localfdr}
\end{equation}
with $f(x) = \eta_0 f_0(x) + (1-\eta_0) f_A(x)$.
As with Fdr, to ensure that the local FDR is increasing with $x$ the density $f_A(x)$
is assumed to be monotonically decreasing. 
The local FNDR is the probability of the alternative under
the observed data, and is thus is given by
$$
\text{fndr}(x) = 1- \fdr(x) \,.
$$

There is also a direct relationship between fdr and Fdr. As can be seen
from its definition Fdr is a conditional average of fdr. Hence, for monotonic
$f_A(x)$  we find $\Fdr(x) \leq \fdr(x)$ for all $x$ \citep{Efr08a}.

Like the HC objective function, fdr and Fdr are  scalars 
and thus transform under a change of coordinates from $x$ to $y$ as
$\fdr(y) = \fdr(x(y))$ and $\Fdr(y) = \Fdr(x(y))$.

\subsection{Signal identification with FDR and FNDR}

A standard approach to obtain a decision
threshold 
with FDR is to refer to the rule of \citet{BH95}
with a cutoff such as $\widehat{\Fdr}(p_{(i)})  \leq 0.05$.
Alternatively, a threshold may be found by controlling 
local FDR, for instance by requiring $\widehat{\fdr}(p_{(i)})  \leq 0.2$
\citep[e.g.][]{Efr08a}.
This ensures that the identified features
are mostly from the alternative  with only little contamination by
unwanted null features.
Conversely, if the interest is to identify true null features then similar
thresholds may be imposed on FNDR rather than FDR \citep{AS2010}.

This illustrated for local FDR and local FNDR  in \figcite{fig:thresh}{\bf b} where
the  signal space is divided by the decision thresholds $x^{\fdr}$ and $x^{\fndr}$ into three
distinct zones  corresponding to areas where one is very
sure about membership to the null (local FNDR $<$ 0.2 or local FDR $>$ 0.8) or to the alternative
(local FDR $<$ 0.2) and one additional intermediate region.

From a classification perspective there exists another threshold ---
the class boundary $x^{\CB}$ --- that provides a natural separation
between null and non-null components. At $x^{\CB}$
the probabilities of membership to the alternative and to the null are equal to $1/2$.
Hence, in terms of local FDR we have  
\begin{equation*}
\fdr(x^{\CB}) = \text{fndr}(x^{\CB}) = \frac{1}{2} \,.
\end{equation*}
As can be seen in \figcite{fig:thresh}{\bf b} by construction $x^{\CB}$
is located inbetween $x^{\fndr}$ and $x^{\fdr}$.
From the definition $\fdr(x^{\CB}) = 1/2$  and \eqcite{eq:localfdr} we obtain the condition 
\begin{equation}
\label{eq:cbthresh}
\eta_0 f_0(x^{\CB}) = (1-\eta_0) f_A(x^{\CB})
\end{equation}
for the CB threshold.

\section{Comparison of CB and HC decision thresholds}

It is now instructive to study the mutual connections among the
various decision thresholds, in particular among
$x^{\HC}$,
$x^{\KS}$,
and $x^{\CB}$.

\subsection{Kolmogorov-Smirnov (KS) decision threshold}

The location $x^{\KS}$ where
 the Kolmogorov-Smirnov objective function
$ | F_A(x) - F_0(x)|$ is maximized is given by
\begin{equation}
\label{eq:ksthresh}
f_0(x^{\KS}) =  f_A(x^{\KS}) \,.
\end{equation}
Thus, the KS decision threshold coincides with the class boundary  $x^{\CB}$
if $\eta_0 = 1/2$.  Thus,
the KS threshold implicitly assumes that
null and non-null components have the same prior probability.

\subsection{HC decision threshold}
\label{sec:hcthresh}

Using \eqcite{eq:hc2-popfunc} we may determine
the population decision threshold
that one tries to estimate by maximizing the empirical HC objective
$\widehat{HC}(x)$.
This leads to 
the general condition 
\begin{equation}
\begin{split}
f_0 \, \left\{ 2 F(1-F) + (F_A-F_0) (1-2 F) \, \eta_0 \right\} & =  \\
f_A \, \{ 2 F(1-F) - (F_A-F_0) (1-2 F) \, (1-\eta_0) \} 
\end{split}
\label{eq:hcthresh}
\end{equation}
that must be satisfied by
the HC decision threshold $x^{\HC}$
(note that in \eqcite{eq:hcthresh} the arguments to $F$, $F_0$ and $F_A$ have been left out for
the sake of clarity).

There are two cases when the HC threshold condition simplifies substantially.
First, if the null and alternative components are well separated:
then $F_A(x^{\HC})= 1$ and
$F_0(x^{\HC}) = 0$ and consequently $F(x^{\HC}) = 1-\eta_0$ so
that \eqcite{eq:hcthresh} reduces to
$$
\eta_0 f_0(x^{\HC}) = (1-\eta_0) f_A(x^{\HC}) \,.
$$
Thus, for well-separated null and alternative
the HC threshold is identical to the CB threshold.

Second, if null and alternative components are very close:
then $F_A(x^{\HC}) \approx F_0(x^{\HC}) $
and \eqcite{eq:hcthresh} becomes
$$
f_0(x^{\HC}) =  f_A(x^{\HC}) \,,
$$
i.e. the HC threshold becomes identical to the KS threshold.

Hence, the HC threshold may be viewed as a compromise between the CB
threshold and the KS threshold. This is directly observed in 
the study of the ``rare-weak'' model (cf. \tabcite{tab:results}{\bf a}).

\section{Rare weak model}

The use of ``Higher Criticism'' is particularly advocated
in settings where the signal is sparse and weak. 
This situation is described by the so-called ``rare weak'' (RW) model
that has been used to study the performance of HC.
In the following we introduce the
RW  model and compare corresponding decision thresholds.

\subsection{Setup of RW model}

The RW model is a  sparse normal mean mixture model
with 
\begin{equation}
\label{eq:rwmodel}
Z \sim (1-\epsilon) N(0, 1) + \epsilon N(\tau, 1) \,.
\end{equation}
Its two parameters $\tau \in [0; \infty]$ and $\epsilon \in [0;1]$ describe
intensity and  sparsity of the signal. 
If $\epsilon$ is small then the non-null features are rare, and likewise if
$\tau$ is small then the effect size is weak (hence the name of the model).
From this mixture we observe $z$-scores $z_1, \ldots, z_d$,
which provide the data from which decision thresholds are inferred.

Despite its simplicity, this model is sufficiently rich to study the
behavior of signal detection and signal identification methods  
\citep{Ing1999,DJ2004,DJ08,DJ2009,XCL2011,JJ2012}.
A generalized RW model with an additional variance parameter in the 
alternative is discussed in \citet{CJJ2011}.

A typical scenario where the RW model naturally arises is in
classification.  For example  consider a two class setting with means
$\bmu_1 =\bmu$ and $\bmu_2=-\bmu$ 
where $\bmu=(\ldots, \mu_0, \ldots, 0, \ldots)^T$ 
is  a $d$-dimensional vector containing either 0 
or $\mu_0$ as components  and with $\epsilon$ describing
the proportion of non-zero entries. Further
assume an identity covariance $\bI_d$ and equal number
of observations $n_1=n_2=n/2$ from the two classes.
Then the corresponding $z$-score vector $(1/n_1+ 1/n_2)^{-1/2} ( \hat\bmu_1 - \hat\bmu_2)$
used for variable selection \citep[e.g.,][]{ZS09} simplifies to $\bz = \sqrt{n} \hat\bmu$.
The $d$ components of  $\bz$ follow the RW model of \eqcite{eq:rwmodel} with
$\tau = \sqrt{n} \mu_0$.  Note the confounding of $n$ and $\mu_0$, so a small number of observations $n$
and large  $\mu_0$ gives rise to the same RW model as large sample size and small $\mu_0$.

Instead of $\epsilon$ and $\tau$ it is sometimes convenient to use
the alternative parameterization
$$
\beta_\epsilon = -\log(\epsilon)/\log(d)
$$
and
$$
r_\tau = \left(\frac{\tau^2}{2}\right)/\log(d) \,
$$
with corresponding backtransformations $\epsilon_\beta = d^{-\beta}$
and $\tau_r = \sqrt{2 r \log(d)}$.

The motivation to use $\beta$ instead of $\epsilon$ to measure sparsity is that for $d$ observations
the smallest possible fraction of the alternative is
$1/d$.  The change of variables maps
$\epsilon \in [\frac{1}{d};1] $  to  $\beta \in [0;1]$.   A sparse setting in the RW model is
characterized by $\beta \in [\frac{1}{2}, 1]$ or equivalently $\epsilon < d^{-1/2}$.
Similarly, the alternative intensity parameter is a map of  
$\tau \in [0; \sqrt{2 \log(d)}]$  to $r \in [0;1]$.  As for $d$ observed  $z$-scores
their maximum is  bounded in expectation by $\sqrt{2 \log(d)}$,  a RW model 
with $r > 1$ contains comparatively well-separated null and alternative components
whereas in a model with $r < 1$ the signal is  weak.

\subsection{Decision boundaries for the RW model}

The RW model is simple enough to allow analytical calculations of 
some decision boundaries.

Using the null and alternative
densities $f_0(z) = \frac{1}{\sqrt{2 \pi} } e^{-z^2/2} $
and $f_A(z) = \frac{1}{\sqrt{2 \pi} } e^{-(z-\tau)^2/2}$ and
distribution functions  $F_0(z) = \Phi(z)$ and
$F_A(z) = \Phi(z-\tau)$ the
 KS decision threshold (\eqcite{eq:ksthresh}) for the RW model is
$$
z^{\KS} = \frac{\tau}{2} \, .
$$

Similarly, the classification class boundary  (\eqcite{eq:cbthresh}) simplifies for
the RW model to
$$
z^{\CB} = \frac{\tau}{2} + \frac{1}{\tau} \log\left(\frac{1-\epsilon}{\epsilon}\right)\, .
$$
For $\epsilon=1/2$ the CB threshold reduces to the KS threshold
and for   $\epsilon \leq 1/2$ we have $z^{\CB} \geq z^{KS}$.
For fixed $\epsilon$ and the effect size $\tau$ large enough the second term above also
vanishes and hence also leads to the KS threshold.
As the 
proportion of non-null features becomes smaller ($\epsilon \rightarrow 0$)
the decision threshold moves to infinity ($ z^{\CB} \rightarrow \infty$).
Thus, if $\epsilon = 0$  no feature will be classified as non-null.

For the HC decision threshold unfortunately no  analytic expression
for $z^{\HC}$ is available.  
 From the general considerations
above (cf. Section~\ref{sec:hcthresh}) we  know that for larger $\tau$ the
HC threshold approximates the CB threshold, and that both reduce
to  the KS threshold for $\epsilon=1/2$.
Furthermore, \citet[][Appendix Eq. 1.1]{DJ2009} show that for the RW model
$\fdr(z^{\HC}) \geq 1/2$.  This together with the monotonicity
of the local FDR in the RW model  implies  that
$$
z^{\HC} \leq z^{\CB} \, .
$$
Thus, in general using the HC decision threshold causes the inclusion of more features than
using the CB threshold.  

Of particular interest is the behavior of the HC threshold for small values of $\epsilon$.
Specifically, if $\epsilon=0$ and $\tau$ is finite then the HC threshold is
also finite. For example, $\epsilon=0$ and $\tau=2$ leads to  $ z^{\HC} \approx 3.35$,
which is distinctly different from the natural class boundary $ z^{\CB} \rightarrow \infty$.
Thus, by construction the HC criterion (and also the KS threshold) encourages false positives in signal identification.


\begin{table}[p]
\caption{a) Comparison of the KS, HC and CB decision thresholds in the RW model, and
b)~Analysis of four cancer gene expression data sets with shrinkage 
discriminant analysis.}
\label{tab:results}

\begin{minipage}{.45\textwidth}

\vspace{4mm}
\begin{tabular}{lrrr}
\multicolumn{4}{l}{\textbf{a) Comparison of Thresholds}} \\
\toprule
Setting & $z^{\KS}$  & $z^{\HC}$ &  $z^{\CB}$ \\
\midrule
\midrule
$\tau=2$  \\
$\epsilon = 0$     & 1 & 3.3514 & $\infty$ \\
$\epsilon = 0.001$ & 1 & 3.0707 & 4.4534 \\
$\epsilon = 0.01$  & 1 & 2.5203 & 3.2976 \\
$\epsilon = 0.1$   & 1 & 1.7574 & 2.0986 \\
$\epsilon = 0.5^*$   & 1 & 1.0000 & 1 \\
\midrule
$\tau=4$  \\  
$\epsilon = 0$     & 2 & 3.3514 & $\infty$ \\
$\epsilon = 0.001^*$ & 2 & 3.6377 & 3.7267\\
$\epsilon = 0.01^*$  & 2 & 3.0965 & 3.1488\\
$\epsilon = 0.1^*$   & 2 & 2.5268 & 2.5493\\
$\epsilon = 0.5^*$   & 2 & 2.0000 & 2 \\
\midrule
$\tau=6$  \\
$\epsilon = 0$     & 3 & 8.1607 & $\infty$ \\
$\epsilon = 0.001^*$ & 3 & 4.1454 & 4.1511 \\
$\epsilon = 0.01^*$  & 3 & 3.7631 & 3.7659 \\
$\epsilon = 0.1^*$   & 3 & 3.3652 & 3.3662 \\
$\epsilon = 0.5^*$   & 3 & 3.0000 & 3 \\
\bottomrule 
\end{tabular}\\
$^*$ Signal identification is  possible as $\epsilon \geq \exp( -\tau^2 /2)$, see Section~\ref{sec:phasespace}.

\end{minipage}
\hspace{5mm} 
\begin{minipage}{.45\textwidth}

\begin{tabular}{lrrr}
\multicolumn{4}{l}{\textbf{b) Cancer Gene Expression Data}} \\
\toprule
Data /  &  \multicolumn{2}{c}{Prediction Error} & Selected \\
Method  &        & & Variables \\
\midrule
\midrule
\multicolumn{4}{l}{ {\bf Prostate} ($d=6033,n=102, K=2$) } \\
CB      & 0.0637 & (0.0053) & 115  \\
HC         & 0.0497 & (0.0045) & 116 \\
FNDR       & 0.0550 & (0.0048) & 131 \\
\midrule
\multicolumn{4}{l}{ {\bf Lymphoma } ($d=4026,n=62, K=3$) }\\
CB      & 0.0211 &(0.0042) & 178 \\
HC         & 0.0000 &(0.0000) & 345 \\
FNDR       & 0.0036 &(0.0018) & 392 \\
\midrule
\multicolumn{4}{l}{ {\bf SRBCT} ($d=2308,n=63, K=4$) } \\
CB     & 0.0000 &(0.0000) & 88 \\
HC         & 0.0007 &(0.0007) & 174 \\
FNDR       & 0.0000 &(0.0000) & 89  \\
\midrule
\multicolumn{4}{l}{ {\bf Brain} ($d=5597,n=42, K=5$) } \\
CB     & 0.1633 &(0.0120) & 78 \\
HC         & 0.1417 &(0.0108) & 131 \\
FNDR       & 0.1525 &(0.0120) & 102 \\
\bottomrule 
\end{tabular}\\
$K$: number of classes in the response variable.

\end{minipage}

\end{table}


A comparison of the KS, HC, and CB thresholds
for some  settings of $\epsilon$ and $\tau$
is given in \tabcite{tab:results}{\bf a}. 
As expected, with increasing $\tau$ the HC and
CB thresholds become very similar and for
$\epsilon = 1/2$ both HC and the CB threshold
reduce to the KS threshold.  Thus, the pattern confirms
the  general relationships of these decision
thresholds discussed above.

In addition, in the RW model 
there exist a further close link between the HC and CB thresholds.
This results from the special structure of the parameter space of
the RW model discussed next.

\subsection{Phase space of the  RW model}
\label{sec:phasespace}

\begin{figure}[thp!]
\begin{center}
\centerline{\includegraphics[width=1\textwidth]{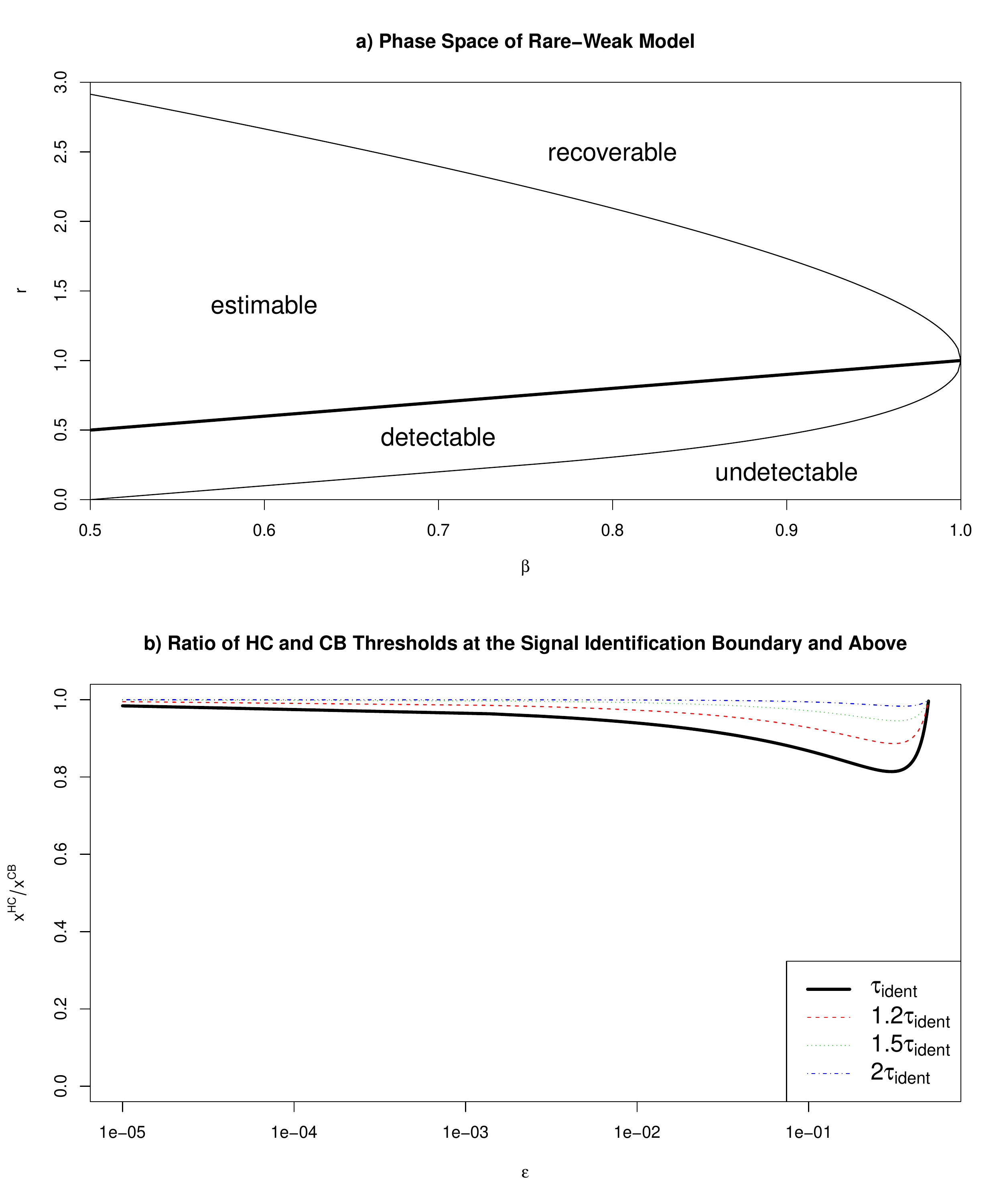}}
\caption{a) Phase space of the RW model following \citet{XCL2011} and \citet{JJ2012}. The bold line shows the
signal identification boundary $r_\text{ident}(\beta) = \beta$ above which signal identification is possible. For details on the four regions see the description in
Section~\ref{sec:phasespace}. 
b) Ratio of $x^\HC$ and $x^\CB$ thresholds  at the signal identification
boundary  (solid line) and above (dotted lines). Note that $\tau_{\text{ident}}(\epsilon) = \sqrt{-2 \log(\epsilon)}$.
\label{fig:phase-ratio}}
\end{center}
\end{figure}

Within the RW model the behavior of signal detection and identification
procedures have been studied extensively. This has
lead to the remarkable insight that there exist several fundamental boundaries in 
its phase space that give rise to four distinct regions, as
illustrated in \figcite{fig:phase-ratio}{\bf a}.

\citet{Ing1999} discovered the \emph{detection boundary}
$$
r_\text{detect}(\beta) = 
\begin{cases}
\beta - \frac{1}{2} & \beta \in [\frac{1}{2}; \frac{3}{4}] \\
 ( 1-\sqrt{1-\beta} )^2 & \beta \in [\frac{3}{4}; 1] \,.
\end{cases}
$$
Below this boundary lies the ``undetectable'' region in which
even signal detection is impossible, i.e.
no method is able to decide whether  $\epsilon \neq 0$.
Conversely, above the detection boundary  it is  possible to consistently 
estimate $\epsilon$ \citep{CJL2007}.

\citet{DJ2004} report the \emph{identification boundary} 
$$
r_\text{ident}(\beta) = \beta \,.
$$
It is only above this boundary in the ``estimable'' and ``recoverable''
regions that  signal identification
by thresholding is actually possible.
In terms of original parameters this corresponds to the conditions
$\tau  \geq  \sqrt{-2 \log(\epsilon)}$
or
$\epsilon \geq \exp( -\tau^2 /2)$.
Directly below this boundary lies the ``detectable'' region where
detection of a signal is possible but not identification.
This shows that signal identification is more difficult 
than signal detection.

Finally, \citet{XCL2011} and \citet{JJ2012} demonstrated the existence of
the \emph{recovery boundary} 
$$
r_\text{recov}(\beta) = ( 1+\sqrt{1-\beta} )^2
$$
above which in the ``recoverable'' region almost all signal can be completely identified.

\subsection{HC threshold as approximation of the natural class boundary}

When comparing the KS, HC and CB decision thresholds in \tabcite{tab:results}{\bf a}
a striking phenomenon can be observed:
whenever signal identification is possible, i.e. if $\epsilon \geq \exp( -\tau^2 /2)$,
then then $z^{\CB}$ and $z^{\HC}$ are very similar.

To investigate this further we computed the ratio of the HC and CB threshold
directly at the signal identification boundary, and above (\figcite{fig:phase-ratio}{\bf b}).
Already at the boundary this ratio is close to 1, especially for small values of $\epsilon$. 
Moving further into the ``estimable'' and ``recoverable'' regions the differences
between the two thresholds become negligible.

Hence, in the RW model in the area where signal identification is possible
$z^{\HC}$ and $z^{\CB}$
are in the worst case very similar and
mostly indistinguishable for practical purposes.

\section{Data examples}

To further study the relationship among the HC, CB, and FNDR decision thresholds 
we analyzed both simulated as well as experimental data.

\subsection{Synthetic data}

\begin{figure}[t]
\begin{center}
\centerline{\includegraphics[width=1\textwidth]{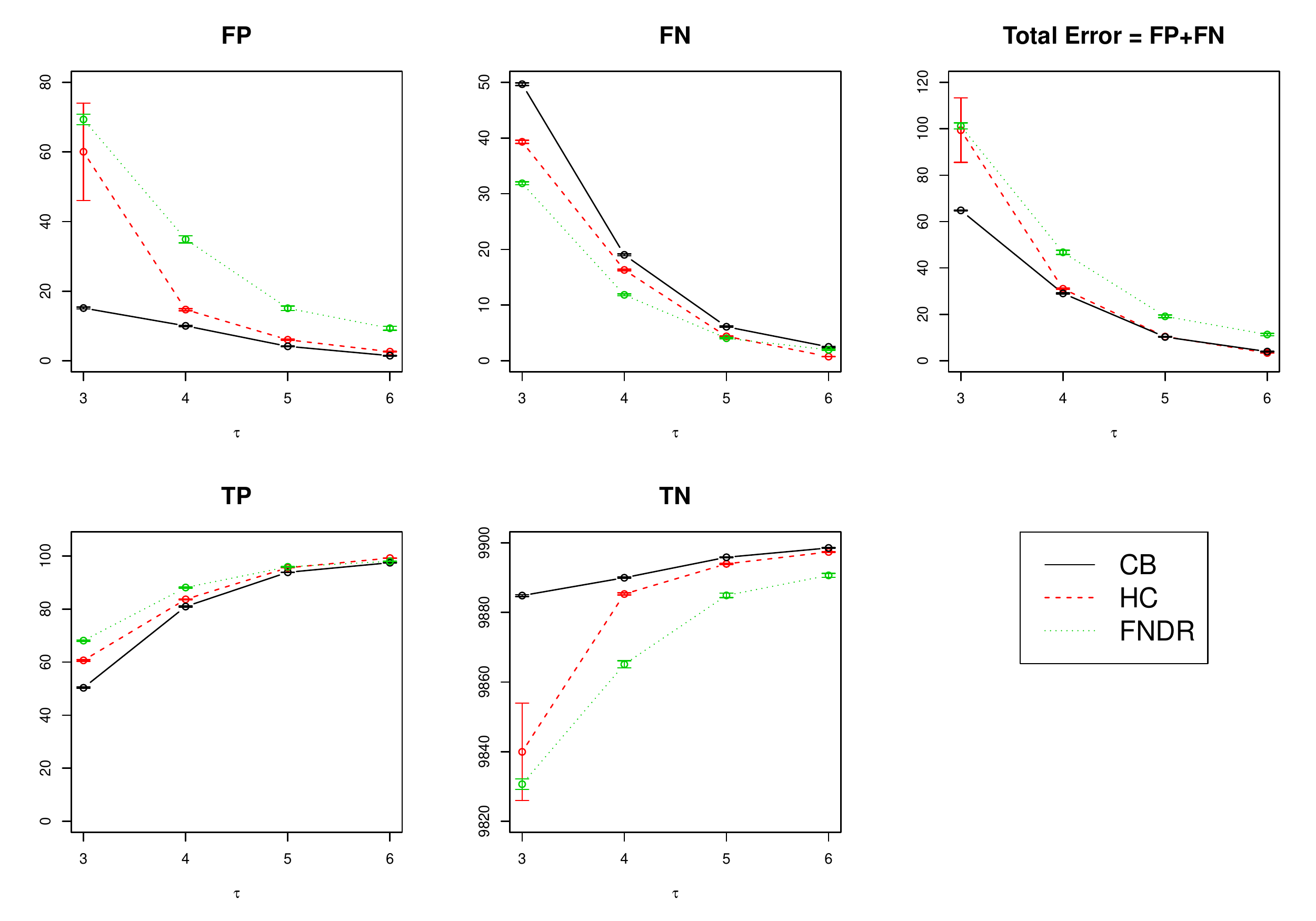}}
\caption{Comparison of errors when
using the HC, CB, and FNDR decision thresholds on data simulated from 
the RW model located directly at the detection boundary ($\epsilon=0.01$ and $\tau=3$)
and above ($\tau>3$). 
\label{fig:simulation}}
\end{center}
\end{figure}

We simulated data from the RW model at the signal identification boundary and above, as follows:
\begin{enumerate}
\item We sampled $d=10,000$ $z$-scores
from the mixture model
\eqcite{eq:rwmodel} with $\epsilon=0.01$ and $\tau \in  \{3,4,5,6\}$.
For $\tau = 3$ this is is a sparse and weak scenario located directly at
 the signal identification boundary ($\epsilon \approx \exp(-\tau^2/2)$).
\item 
From the test statistics  $z_1, \ldots, z_d$ we computed $p$-values according to
$p_i = 1-F_0(z_i)$.
\item Subsequently, the  empirical HC threshold
was obtained by maximization of \eqcite{eq:hcfunc}.
\item In addition, local FDR was estimated using the {\tt fdrtool} algorithm \citep{Str08b,Str08c} and correspondingly the CB (local FDR = 0.5) and FNDR (local FDR = 0.8)
decision thresholds were identified.
\item For each of the three investigated thresholds (HC, CB, FNDR) the
number of false positives (FP), false negatives (FN), true positives (TP) and
true negatives (TN) were determined.
\item The simulations were repeated $B=1000$ times to estimate mean errors and their
standard deviations.
\end{enumerate}

The results are visualized in \figcite{fig:simulation}.
As expected, the HC and CB thresholds yield  similar results with
growing  $\tau$.   However, if the signal is weak (small $\tau$)
signal identification with HC leads to many more  more false positives,
and in addition the variability of the error rates for HC is very large.
Conversely, in this situation the CB threshold is more cautious and
thus results in more false negatives.  For all settings the error rates of HC
are found in between those of CB and FNDR.
Interestingly, the total error (FP+FN) 
 is smallest when using the CB threshold.

We also repeated this study with other sparsity settings 
$\epsilon > 0.01$.  The resulting error plots all show exactly the same pattern
of convergence of the CB and and HC methods as \figcite{fig:simulation}.

\subsection{Gene expression data}

Next, we also analyzed four clinical
gene expression data
sets related to prostate cancer \citep{SF+02}
lymphoma \citep{AED+00}, small round blue cell tumors (SRBCT)
 \citep{KW+01}, and brain cancer \citep{PTG+02}.
Previously, in \citet{AS2010} we have compared
the relative effectiveness of the FNDR and HC thresholds
to select relevant genes in shrinkage discriminant analysis
using CAT scores \citep{ZS09}.   

In \tabcite{tab:results}{\bf b}
we show in addition the estimated prediction error and the 
number of selected variables for the CB threshold.
Generally, using the CB decision threshold leads to
the smallest predictor sets. Except for the prostate data  the
number of selected genes is roughly  half compared to using 
the HC threshold
as criterion.  As the predictor error is only slightly increased
 we conclude that most of the additionally included predictors 
by HC are false positives.

For practical analysis of gene expression data this implies that
using $x^{\CB}$ yields --- in comparison with $x^{\HC}$ --- smaller and hence more interpretable
predictor
gene sets without compromising  prediction error.

\section{Discussion}

Our investigation of the relationship of the HC and FDR methods 
started with the aim to better understand HC as
a method for signal identification.
In the context of variable selection for classification \citet{DJ08} 
demonstrated empirically that using $x^{\HC}$ as a decision threshold
outperforms competing procedures, in particular those using a threshold based on
FDR. \citet{DJ2009} further justified HC as a signal identification
procedure by showing that $x^{\HC}$
minimizes an approximation to the missclassification error.

Here, we argue that the HC decision threshold may also be viewed as an
approximation of the natural class boundary between the null and
alternative groups in the RW mixture model.  This CB threshold can
be directly expressed in terms of local FDR and local FNDR.
Importantly, in the RW model in the region
of the phase space where signal identification is possible both
thresholds are either very similar or practically indistinguishable.
Interestingly, computing this threshold via HC uses only
distribution functions ($F$, $F_0$, and $F_A$, cf.  \eqcite{eq:hc2-popfunc}) but
in addition requires optimization, whereas computation via local FDR is direct 
but employs densities ($f$, $f_0$, and $f_A$, cf.  \eqcite{eq:localfdr}) which are more difficult to obtain.

If the two thresholds are notably different then 
using the HC threshold leads to the inclusion
 of more false positives, and conversely
the CB threshold yields a more compact feature set 
but with slightly increased prediction error. 
In short, the CB threshold is more cautious
than the  HC threshold (and the FNDR threshold).

Hence, our study provides further support to the excellent performance
of HC for signal identification.  However, our conclusions 
are different from that of \citet{DJ08,DJ2009}.
First, we show that false discovery rates, properly applied, are 
indeed perfectly useful for signal identification, which has been disputed earlier.
Second, the convergence of the CB and HC thresholds in the ``estimable''
and ``recoverable'' regions indicates that this is what HC is actually 
approximating.  

In general, estimation of the CB threshold is a challenging problem as this requires the fit of 
a mixture model and estimation of the mixing density.  In contrast, the empirical
HC threshold can readily be determined using $p$-values computed from $F_0$ alone. Thus, for signal identification the HC approach provides a simple yet effective means to 
approximate the CB threshold.

\section*{Acknowledgements}

Part of this work was supported by BMBF grant  no. 0315452A (HaematoSys project).
The authors would like to thank the anonymous referees for their very valuable comments
and suggestions.


\bibliographystyle{apalike}
\bibliography{preamble,econ,genome,stats,array,sysbio,misc,molevol,med,entropy}

\newcommand{\noopsort}[1]{} \newcommand{\printfirst}[2]{#1}
  \newcommand{\singleletter}[1]{#1} \newcommand{\switchargs}[2]{#2#1}
\begin{thebibliography}{}

\bibitem[Ahdesm\"aki and Strimmer, 2010]{AS2010}
Ahdesm\"aki, M. and Strimmer, K. (2010).
\newblock Feature selection in omics prediction problems using cat scores and
  false non-discovery rate control.
\newblock {\em Ann. Appl. Statist.}, 4:503--519.

\bibitem[Alizadeh et~al., 2000]{AED+00}
Alizadeh, A.~A., Eisen, M.~B., Davis, R.~E., Ma, C., Lossos, I.~S., Rosenwald,
  A., Boldrick, J.~C., Sabet, H., Tran, T., Yu, X., Powell, J.~I., Yang, L.,
  Marti, G.~E., Moore, T., Hudson, J., Lu, L., Lewis, D.~B., Tibshirani, R.,
  Sherlock, G., Chan, W.~C., Greiner, T.~C., Weisenburger, D.~D., Armitage,
  J.~O., Warnke, R., Levy, R., Wilson, W., Grever, M.~R., Byrd, J.~C.,
  Botstein, D., Brown, P.~O., and Staudt, L.~M. (2000).
\newblock Distinct types of diffuse large {B}-cell lymphoma identified by gene
  expression profiling.
\newblock {\em Nature}, 403:503--511.

\bibitem[Anderson and Darling, 1954]{AndersonDarling1954}
Anderson, T.~W. and Darling, D.~A. (1954).
\newblock A test of goodness of fit.
\newblock {\em J. Amer. Statist. Assoc.}, 49:765--769.

\bibitem[Benjamini and Hochberg, 1995]{BH95}
Benjamini, Y. and Hochberg, Y. (1995).
\newblock Controlling the false discovery rate: a practical and powerful
  approach to multiple testing.
\newblock {\em J. R. Statist. Soc. B}, 57:289--300.

\bibitem[Cai et~al., 2011]{CJJ2011}
Cai, T.~T., Jeng, X.~J., and Jin, J. (2011).
\newblock Optimal detection of heterogeneous and heteroscedastic mixtures.
\newblock {\em J. R. Statist. Soc. B}, 73:629--662.

\bibitem[Cai et~al., 2007]{CJL2007}
Cai, T.~T., Jin, J., and Low, M.~G. (2007).
\newblock Estimation and confidence sets for spare normal mixtures.
\newblock {\em Ann. Statist.}, 35:2421--2449.

\bibitem[Darling, 1957]{Darling1957}
Darling, D.~A. (1957).
\newblock The {Kolmogorov-Smirnov}, {Cram\' er-von Mises} tests.
\newblock {\em Ann. Math. Stat.}, 28:823--838.

\bibitem[Donoho and Jin, 2004]{DJ2004}
Donoho, D. and Jin, J. (2004).
\newblock Higher criticism for detecting sparse heterogeneous mixtures.
\newblock {\em Annals of Statistics}, 32:962--994.

\bibitem[Donoho and Jin, 2008]{DJ08}
Donoho, D. and Jin, J. (2008).
\newblock Higher criticism thresholding: optimal feature selection when useful
  features are rare and weak.
\newblock {\em Proc. Natl. Acad. Sci. USA}, 105:14790--15795.

\bibitem[Donoho and Jin, 2009]{DJ2009}
Donoho, D. and Jin, J. (2009).
\newblock Feature selection by higher criticism thresholding achieves the
  optimal phase diagram.
\newblock {\em Phil. Trans. R. Soc. A}, 367:4449--4470.

\bibitem[Efron, 2008]{Efr08a}
Efron, B. (2008).
\newblock Microarrays, empirical {Bayes}, and the two-groups model.
\newblock {\em Statist. Sci.}, 23:1--22.

\bibitem[Genovese and Wassermann, 2002]{GW02}
Genovese, C. and Wassermann, L. (2002).
\newblock Operating characteristics and extensions of the false discovery rate
  procedure.
\newblock {\em J. R. Statist. Soc. B}, 64:499--517.

\bibitem[Ingster, 1999]{Ing1999}
Ingster, Y.~I. (1999).
\newblock Minimax detection of a signal for $l^p_n$ balls.
\newblock {\em Math. Methods. Statist.}, 7:401--428.

\bibitem[Jager and Wellner, 2007]{JW2007}
Jager, L. and Wellner, J.~A. (2007).
\newblock Goodness-of-fit tests via phi-divergences.
\newblock {\em Ann. Statist.}, 35:2018--2053.

\bibitem[Ji and Jin, 2012]{JJ2012}
Ji, P. and Jin, J. (2012).
\newblock {UPS} delivers optimal phase diagram in high-dimensional variable
  selection.
\newblock {\em Ann. Statist.}, 40:73--103.

\bibitem[Khan et~al., 2001]{KW+01}
Khan, J., Wei, J.~S., Ringner, M., Saal, L.~H., Ladanyi, M., Westermann, F.,
  Berthold, F., Schwab, M., Antonescu, C.~R., Peterson, C., and Meltzer, P.~S.
  (2001).
\newblock Classification and diagnostic prediction of cancers using gene
  expression profiling and artificial neural networks.
\newblock {\em Nature Med.}, 7:673--679.

\bibitem[Langaas et~al., 2005]{LLF05}
Langaas, M., Lindqvist, B.~H., and Ferkingstad, E. (2005).
\newblock Estimating the proportion of true null hypotheses, with application
  to {DNA} microarray data.
\newblock {\em J. R. Statist. Soc. B}, 67:565--572.

\bibitem[Pomeroy et~al., 2002]{PTG+02}
Pomeroy, S.~L., Tamayo, P., Gaasenbeek, M., Sturla, L.~M., Angelo, M.,
  McLaughlin, M.~E., Kim, J. Y.~H., Goumnerova, L.~C., Black, P.~M., Lau, C.,
  Allen, J.~C., Zagzag, D., Olson, J.~M., Curran, T., Wetmore, C., Biegel,
  J.~A., Poggio, T., Mukherjee, S., Rifkin, R., Califano, A., Stolovitzky, G.,
  Louis, D.~N., Mesirov, J.~P., Lander, E.~S., and Golub, T.~R. (2002).
\newblock Prediction of central nervous system embryonal tumour outcome based
  on gene expression.
\newblock {\em Nature}, 415:436--442.

\bibitem[Singh et~al., 2002]{SF+02}
Singh, D., Febbo, P.~G., Ross, K., Jackson, D.~G., Manola, J., Ladd, C.,
  Tamayo, P., Renshaw, A.~A., D'Amico, A.~V., Richie, J.~P., Lander, E.~S.,
  Loda, M., Kantoff, P.~W., Golub, T.~R., and Sellers, W.~R. (2002).
\newblock Gene expression correlates of clinical prostate cancer behavior.
\newblock {\em Cancer Cell}, 1:203--209.

\bibitem[Strimmer, 2008a]{Str08b}
Strimmer, K. (2008a).
\newblock {fdrtool}: a versatile {R} package for estimating local and tail
  area-based false discovery rates.
\newblock {\em Bioinformatics}, 24:1461--1462.

\bibitem[Strimmer, 2008b]{Str08c}
Strimmer, K. (2008b).
\newblock A unified approach to false discovery rate estimation.
\newblock {\em BMC Bioinformatics}, 9:303.

\bibitem[Tukey, 1976]{Tukey1976}
Tukey, J.~W. (1976).
\newblock {T13 N: the higher criticism}.
\newblock {Course Notes, Statistics 411}, Princeton Univ.

\bibitem[Xie et~al., 2011]{XCL2011}
Xie, J., Cai, T.~T., and Li, H. (2011).
\newblock Sample size and power analysis for sparse signal recovery in
  genome-wide association studies.
\newblock {\em Biometrika}, 98:273--290.

\bibitem[Zuber and Strimmer, 2009]{ZS09}
Zuber, V. and Strimmer, K. (2009).
\newblock Gene ranking and biomarker discovery under correlation.
\newblock {\em Bioinformatics}, 25:2700--2707.

\end{thebibliography}

\end{document}